# Development of a Multi-Task Learning V-Net for Pulmonary Lobar Segmentation on Computed Tomography and Application to Diseased Lungs

*Marc Boubnovski Martell[1], Mitchell Chen[1,3], Kristofer Linton-Reid[1], Joram M. Posma[2], Susan J Copley[1,3], Eric O. Aboagye[1]\**

[1]*Comprehensive Cancer Imaging Centre, Department of Surgery & Cancer, Imperial College London, Hammersmith Hospital, London W12 0NN, UK;* [2]*Department of Metabolism, Digestion and Reproduction, South Kensington, London SW7 2AZ, United Kingdom; and* [3]*Department of Radiology, Hammersmith Hospital, Imperial College NHS Trust, London W12 0HS, United Kingdom.*

*Marc Boubnovski Martell – first author; \* Corresponding author, eric.aboagye@imperial.ac.uk*

*Funding: The work was funded by Stratified Medicine Graduate Training Programme (STRATiGRAD) award WROC_I36048, the Imperial College NIHR Biomedical Research Centre, the Imperial College Cancer Research UK Centre, and the Imperial College Experimental Cancers Centre.*

*Abstract*
Automated lobar segmentation allows regional evaluation of lung disease and is important for diagnosis and therapy planning. Advanced statistical workflows permitting such evaluation is a needed area within respiratory medicine, their adoption remains slow, with poor workflow accuracy. Diseased lung regions often produce high-density zones on CT images, limiting an algorithm's performance. This can result in a failure to accurately discriminate physically damaged lobes, due to oblique or anatomically incomplete fissures. This motivated the development of an improved machine learning method to segment lung lobes while retaining high performance in mild and severe disease alike. We present new methodology that utilises tracheobronchial tree information, to enhance segmentation accuracy through the algorithm's spatial familiarity to more accurately define lobar extent. The method undertakes parallel segmentation of lobes and auxiliary tissues simultaneously by employing a multi-task learning (MTL) in conjunction with V-Net-attention, a popular convolutional neural network in the imaging realm. In keeping with the model's adeptness for better generalisation, high performance was retained in an external dataset of patients with four distinct diseases: severe lung cancer, COVID-19 pneumonitis, collapsed lungs and Chronic Obstructive Pulmonary Disease (COPD), even though the training data included none of these cases. The benefit of our external validation test is specifically relevant since our choice includes those patients who have diagnosed lung disease with associated radiological abnormalities. To ensure equal rank is given to all segmentations in the main task we report the following performance (Dice score) on a per-segment basis: normal lungs 0.97, COPD 0.94, lung cancer 0.94, COVID-19 pneumonitis 0.94 and collapsed lung 0.92, all at $p<0.05$. The model was skilled at segmenting lobes and boundaries well, despite the vast impact of disease, showing that the associated tissue learning, serves to improve overall segmentation accuracy. Notably, even when segmenting lobes with extensive structural abnormality, the model managed to maintain state-of-the-art accuracy. The proposed model can therefore be readily adopted in the clinical setting as a robust tool for radiologists and researchers to define lobar distribution of lung disease, and as a basis for therapeutic intervention.

1. INTRODUCTION

The advancing role of AI methods in analysing complex pulmonary data makes medical imaging a more evidence-based, numerically measurable field (Mansoor et al., 2015). Allowing clinicians to evaluate quantified data from a variety of organs and abnormalities within the respiratory system is vital for multiple clinical applications and becoming ubiquitous in pulmonary analysis (Ng et al., 2020). Given that in many instances, CT abnormalities may be present before patient symptoms occur, this modality's role is viewed as a very sensitive imaging technique vital for detailed pulmonary review and precise diagnosis (Shen et al., 2017).

In the last decade, while carrying out automatic/semi-automatic lobe segmentation tasks, researchers typically include handcrafted feature engineering in the design. The issue was that they were often not robust when fissures are incomplete or distorted or adjacent to extensive lung abnormality. Thus, more recent studies began combining supplementary anatomical information such as lung, airways, and vessels along with main lobar data to improve segmentation accuracy (Kuhnigk et al., 2005). Although those methods have achieved some progress, they still face notable problematic issues. Their multi-stage pipelines are cumbersome, demanding ample time to digest long computational sequences; the segmentation accuracy is conditional to the atlas-quality. Standard automated methods also often fail to deliver adequate precision when wide variations in shape or surface of fissures and tissues occur. As such, diseased and anatomically complex tissue present a non-trivial uncertainty for the segmentation task.

Pulmonary analysis with AI is still far from full automation. However, the recent debut of deep learning is viewed to vastly advance the organ segmentation task. Fully convolutional networks (FCN) (Shelhamer et al., 2017) have become a promising technology with their ability to solve computational analysis, bypassing costly and lengthy expert-done feature engineering. Learning orderly in an automated way from the data, FCNs are adept in objective function optimisation. With FCNs rapidly shifting clinical imaging, multiple manuscripts have offered 2D based analysis which misses the vital contextual knowledge available in the 3D setup. This knowledge is likely to be very decisive for statistical accuracy as the key anatomical familiarity from adjacent tissue slices is helpful in instances where lobar fissures are incomplete. In turn, 3D context may arguably benefit the imaging ambiguity caused by incomplete fissure appearance.

Given that CT measurements hold encoded data in the 3D format, in order to capture this information, an extension from 2D to 3D convolutions is needed. An earlier study by Park et al. (2020) introduced 3D U-Net for automated pulmonary lobe segmentation; V-net (fully regularised) reported by Ferreira et al. (2018); as well as Lee et al's PLS-Net (2019) with asymmetric encoder-decoder configuration -- all three represent relevant examples of advanced work in the segmentation setting. Computational-wise complexity and numerous parameters of 3D networks mean lengthy training sessions. Additionally, the optimisation of million's of parameters demands extensive datasets in order to be done properly. Importantly, even using advanced graphic cards, full volumetric CT datasets is not an option for most networks' architectures. Thus, either the network must be carried out on 3D patches or be applied on downsampled images.

The best results so far achieved in lobar segmentations end-to-end was obtained by Lee et al. (2019). Their proposed PLS-Net, is an asymmetric encoder-decoder architecture, which was specifically adapted to work on the entire volumetric CT. Notably, the authors trained their model on a larger dataset (210 CT scans) than other studies. Thus, it is rather unclear whether their model's novelty or larger dataset size contributed to the better performance. Another consideration in this study would be accuracy when the dose of the CT scan decreases, or interspaced sections are acquired. Since the authors trained the model only on entire CT scans, the performance might notably weaken on low-dose scans or those with interspaced slices, such as COVID-19 scans from MOSMED.

Our work was motivated by two previous studies: by Ferreira et al. (2018) and by Wang et al. (2017). In the former, the authors devised an automated lobes segmentation method by focusing on difficult to define lobar fissures. Given its limited dataset, V-Net yielded accurate results, recognizing the fissures even when boundaries were difficult to distinguish in the images. The latter compared a multi-task FCN to a distinct method, namely multi-class FCN. Both these studies concurrently segmented lungs, heart, and clavicles. The difference is that the multi-task deep neural network links multiple outputs' paths with different objective functions on top of a standard base network. Unlike the multi-class method, it uses a combined objective function for numerous classes. The paper showed that multi-task learning converges faster during training. Moreover, it has faster test time speed, furnishing superior segmentation. Wang et al. (2017) argue that one possible explanation for better results is that the contextual information of the lung helps the network to determine the boundary of the clavicles.

In our study, we proposed a multi-task mode to augment the robustness of pulmonary lobar segmentation. MTL is a deep learning technique that enables neural networks to solve two interrelated tasks simultaneously. This method, known to act as a regulariser, makes the model more robust as it forces the network to perform well on two different interrelated tasks (Ruder et al., 2017). This view is based on the notion that incorporating the auxiliary segmentation of the trachea and bronchi into principal segmentation of five lung lobes improves the model's accuracy given that the pulmonary system contains these anatomically associated central structures (He et al., 2019).

In cases where the patients' lungs are diseased or demonstrate abnormalities, we expect our model to use the position of the trachea/bronchi to help it recognise the spatial environment of the lobes. The underlying premise of this approach implies that the shared optimisation of those coupled segmentations should point to more robust results in the context of lobar volumetric awareness. Such technology may become an influential element for Computer-Aided Diagnosis workflows used in radiology departments (Doel et al., 2015); including further analysis in radiomics workflows. More accurately defined objective volumetric lung lobe data signifies real progress for both lobar disease analysis and subsequent treatment measures.

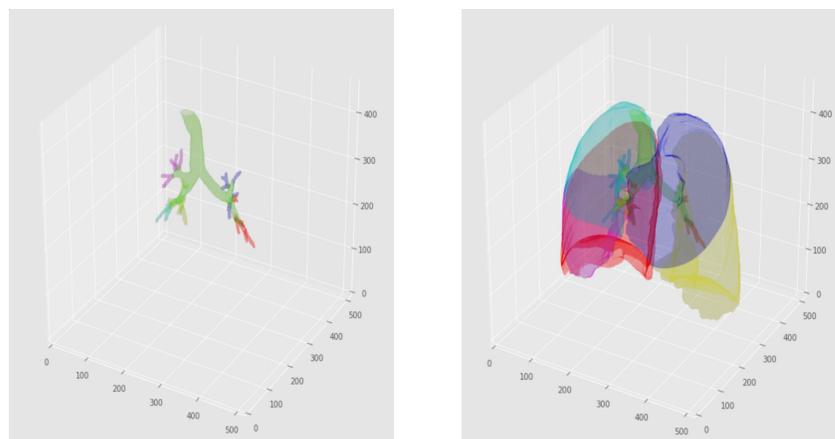

(a) Trachea, lobar and segmental bronchi    (b) Whole lung

Fig. 1. Visualisation of a segmented lung shows trachea, lobes and segmental and lobar bronchi. The image is from a patient in the training dataset.

Our hypothesis states that the inclusion of tracheobronchial tree information in a multi-task deep learning V-Net with attention model improves the accuracy of lobar segmentation of different lung diseases. We created a system that will generalise better for interpreting new datasets. The effectiveness of this segmentation tool was tested on CT scans with Chronic Obstructive Pulmonary Disease (COPD; characterised by bronchitis and emphysema), COVID-19 pneumonitis, collapsed lung, and extensive lung cancer, aiming to assess whether extra information at a lobar level can improve segmentation in these cases. During the COVID-19 pandemic, radiology reporting of Thoracic CTs is put under enormous strain; therefore, such automatic tools could be of value in radiology departments for the objective quantification of disease extent which may be prognostic.

## 2. Materials and Methods

### 2.1. Training dataset

The training dataset consists of 50 thoracic CT scans reported as 'Normal; or 'Near normal' (cases with small nodules but preserved lung architecture were acceptable), and their corresponding segmentation masks (pulmonary lobes, trachea, and bronchi divided by the lobe they are in). Lobar and airway segmentations were prepared and verified by a board-certified clinical radiologist of four years of chest imaging experience, using Pulmonary Toolkit (Doel et al., 2021), an open-source image processing kit that runs in the MATLAB environment (MathWorks, Natick, Massachusetts, USA), and 3D Slicer (3D Slicer) (Fedorov et al, 2012). In our airway segmentation, we have segmented lobar, segmental, and in some cases subsegmental airways where possible, and grouped them by the lobe they are in. 38 of these cases were drawn from the SPIE-AAPM Lung CT Challenge cohort, publicly available from The Cancer Imaging Archive (TCIA) (Armato et al., 2015, Clark et al., 2013). The remaining are 12 fully anonymised studies randomly selected from an ethically approved retrospective local lung cancer study (REC: 18HH4616), of cases diagnosed between 2012-2018 at Imperial College Healthcare NHS Trust, London, UK.

### 2.2. Validation and test dataset

For validation and test, this retrospective study used the largest publicly available, annotated dataset (Tang et al., 2019) consisting of 50 CT scans with their corresponding masks (five lobes, trachea and lobar bronchi) annotated by radiologists using the Chest Image Platform; this was to ensure that our results could be compared with other studies on the same dataset in the future. The data was collated for the LUNA16 competition from a subset of the LIDIC dataset with 1,063 scans (50.7% female; median age = 60.1 years); the inclusion criteria for scans mandated slices of less than 3mm and consistent spacing between slices. The LIDIC images were collected in 2004 using various scanners with diverse technical parameters creating heterogeneous data that were all anonymised to protect-patient health information (Hancock et al., 2016). We split the data randomly (stratified randomisation): 50% validation, and 50% testing. The CT resolution was 512 x 512 x L; L ∈ [113,561] is the number of slices along the long axis of the body.

### 2.3. Population characteristics segmentation

To maintain patient anonymity, all CT scans had fully anonymised DICOM files; thus, no information on the metadata was available. Table I shows the voxel distribution per class in the training and validation datasets (75 CT scans in total), the information on the test set was not included in our model in order to avoid bias.

TABLE I
CT VOXEL PERCENTAGE REPRESENTATION FOR EACH CLASS

| Class | Percentage representation % | | | | |
|---|---|---|---|---|---|
| | Normal Lungs | Collapsed Lung | Cancer | COVID-19 | COPD |
| Background | 88.3 | 89.9 | 88.9 | 88.2 | 88.4 |
| LR lobe | 2.69 | 2.55 | 2.59 | 3.09 | 2.63 |
| MR lobe | 1.07 | 0.44 | 0.9 | 1.03 | 0.98 |
| UR lobe | 2.43 | 1.96 | 2.47 | 2.08 | 2.34 |
| LL lobe | 2.48 | 2.22 | 2.36 | 2.72 | 2.62 |
| UL lobe | 2.86 | 2.82 | 2.68 | 2.78 | 2.89 |
| Trachea & Primary Bronchi | 0.14 | 0.09 | 0.08 | 0.08 | 0.11 |
| Lobar & Segmental Bronchi | 0.03 | 0.02 | 0.02 | 0.02 | 0.03 |

### 2.4. External test set - disease cases

The diseased set consists of four cohorts: COVID-19 pneumonitis, lung cancer, collapsed lung and COPD – comprising six thoracic CT scans each. The COVID-19 cases were selected from the RSNA International COVID-19 Open Radiology Database (RICORD). All patients in this cohort had a positive COVID-PCR result and CT features characteristic of the disease. The lung cancer and collapsed cases were chosen from the local lung cancer study. The extensive lung cancer cases all showed significant architectural distortion secondary to the underlying malignancy. The lung collapse cases showed either a complete or significant partial single lobar collapse. The COPD cases are classified clinically as Global Initiative for Chronic Obstructive Lung Disease (GOLD) stage II-IV (Global Strategy for the Diagnosis, 2018), and manifest on CT as varying degrees of pulmonary emphysema (mean: 13.1, standard deviation: 3.3%), defined as the percentage of voxels with Hounsfield unit of -950HU or less (%LAA-950) (Gevenois et al., 1995).

Lobar and airway segmentations were prepared in the same way as the training dataset. In the disease validation cohorts, solid tumour has been excluded from the segmentation. Collapsed lung has been included where possible, but equivocal areas were excluded where there was radiological uncertainty on the exact tumour border with pulmonary atelectasis.

TABLE II

Percentage of emphysema and emphysema percentile density (HU) in COPD patients

|  | Percentage emphysema (%) | | | | | | Emphysema percentile density (HU) | | | | | |
|---|---|---|---|---|---|---|---|---|---|---|---|---|
|  | Patient 1 | Patient 2 | Patient 3 | Patient 4 | Patient 5 | Patient 6 | Patient 1 | Patient 2 | Patient 3 | Patient 4 | Patient 5 | Patient 6 |
| Both lungs | 11.304 | 11.674 | 13.104 | 9.632 | 20.104 | 12.6 | -936 | -937 | -946 | -932 | -963 | -940 |
| Right lung | 11.947 | 11.333 | 14.261 | 7.045 | 21.918 | 12.69 | -939 | -935 | -948 | -921 | -967 | -941 |
| Left lung | 10.529 | 12.1 | 11.921 | 12.68 | 18.073 | 12.503 | -933 | -939 | -943 | -943 | -958 | -940 |
| RU Lobes | 11.377 | 9.692 | 11.39 | 10.495 | 26.042 | 16.232 | -936 | -929 | -942 | -935 | -976 | -954 |
| RM Lobes | 12.915 | 10.66 | 17.53 | 4.429 | 23.999 | 11.193 | -943 | -933 | -954 | -911 | -971 | -935 |
| RL Lobes | 11.915 | 13.275 | 16.332 | 4.517 | 16.773 | 8.593 | -939 | -942 | -952 | -907 | -955 | -920 |
| LU Lobes | 12.153 | 13.562 | 11.276 | 17.734 | 20.285 | 15.092 | -940 | -945 | -941 | -956 | -963 | -950 |
| LL Lobes | 9.399 | 10.302 | 12.596 | 6.524 | 15.426 | 8.586 | -928 | -929 | -944 | -918 | -951 | -921 |

*2.5. Preprocessing CT scans*

Dealing with high variation in data is a typical job performed during preprocessing. The first step was to truncate all the values of the Hounsfield unit (HU) [between -1000,400]. The HU is a metric that measures the density of materials where air has a value of -1000 HU and bone has a value of 1000 HU, values outside the range of [-1000,400] are not considered necessary for lobar segmentation. We then normalised the CT scans to zero mean with unit variance.

To fit the images into a 3D model, we resized all images, making them of the same size satisfying the memory constraints of the GPU. There are two main approaches to do this that achieve similar performance: downsampling CT images or applying a patch-based method by partitioning the images into overlapping patches. Since we needed to incorporate trachea and bronchial segmentation as an auxiliary task into the segmentation, we opted for the downsampling technique, which ensured that the trachea and bronchi were present in each image. In order to downsample the image, its size was chosen to be the same on all three axes. If one axis is greater resolution than another, the performance along lower resolution axes is worse -- as they contain less information than the other axes. All CT scans (with their masks) were resized to 128x128x128 by using linear and neighbour interpolation, respectively, as it was the largest cube size shape we could fit in memory.

3 Model

*3.1. Model for segmenting lung structures*

In this study, we applied the V-Net architecture using the TensorFlow 1.15 and Keras 2.3 frameworks in Python. The V-Net is an autoencoder and was developed for the purpose of 3D image segmentation (Milletari et al., 2016); its residual connections aid networks to converge and perform well on small datasets. Figure 2 shows the model's architecture, where the input image is size 128x128x128x1 (height, width, depth and channels) and passes through six encoder layers. The output of the main network is 128x128x128x6, where the number of channels correspond to the target classes: five lobes and background and the auxiliary (aux) output is 128x128x128x6 where the classes are the trachea, lobar bronchi and background.

Mitigating the vanishing gradient effect, we selected the Parametric Rectified Linear Unit (PReLU) as the activation function. As suggested by the original V-Net paper (Milletari et al., 2016), we used convolutional layers with a stride of 2x2x2 instead of the max-pooling operation. The last layer before each output was a 1x1x1 convolution with a softmax activation function that gives us the probability that each voxel belongs to one of the classes.

To enhance the model's generalisability, various regularisation techniques were used including dropout, batch normalisation and MTL. Training the model on distinct but relevant tasks has been shown to increase a model's performance (Caruana et al., 1997). The performance of the model is measured using the Dice score, a function of sensitivity and specificity, that is typically used in segmentation studies, permitting a comparison of our model's performance to others. We compared our model to the model that was developed by investigators who released the original dataset. A *P*-value obtained using a t-test smaller than 0.05 between the models was considered statistically significant for independent comparisons.

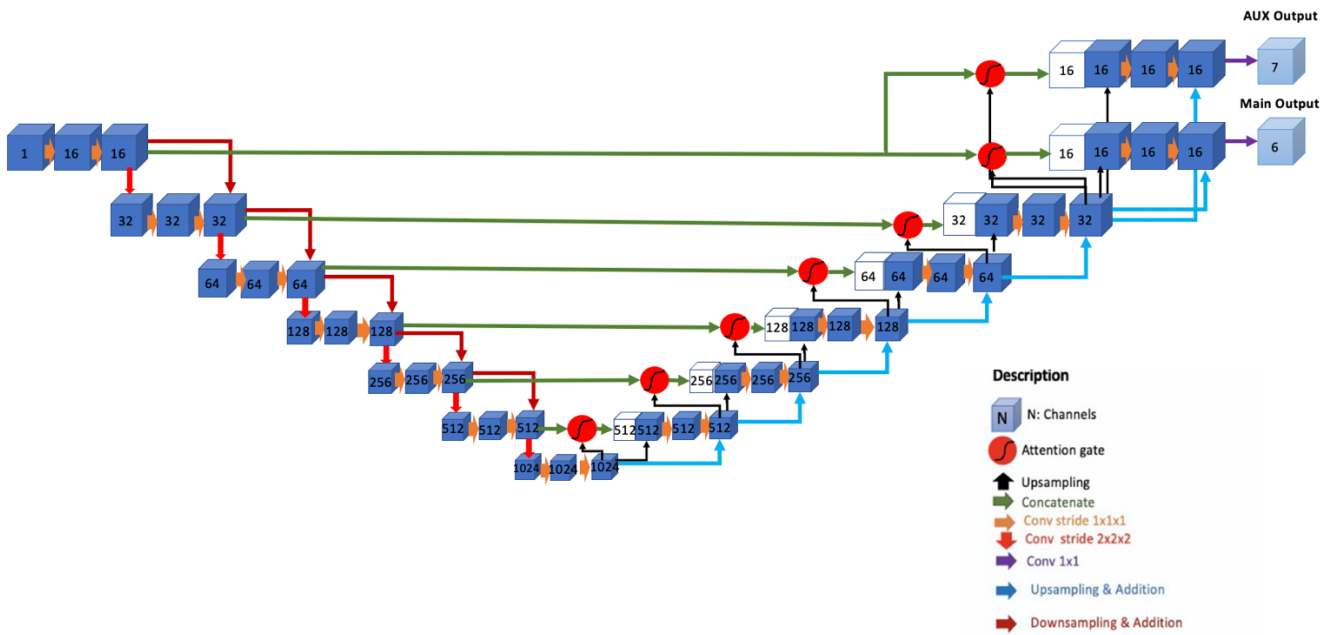

Fig. 2. The neural network architecture for lungs segmentation: V-Net with MTL at the last layer. The network outputs the segmentation of lobes from the main output and the segmentation of bronchi and trachea from the AUX output.

## 3.2. Regularisation

With many parameters in the network and a small training dataset, regularisation is crucial for achieving a robust model that performs well on the test set. Thus, we have incorporated two types of regularisations – dropout and batch normalisation.

## *3.3. Multi-task learning*

MTL is the process when the network is used to solve various tasks simultaneously. Frequently a model struggles to separate important from pointless features due to extreme noise; insufficient sample quantity and dimensionality matters in data. By performing feature selection concurrently for two germane tasks, MTL can aid the model by centring on the most impactful features (Caruana et al., 1997). Training the model on distinct but relevant tasks has been shown to increase model's performance (Wang et al., 2017). Besides, sharing the weights between two tasks, it acts as a regulariser and forces the network to learn a suitable representation to solve both tasks. The reason for choosing to segment bronchi and trachea revolves around extra spatial information they can fill, helping to predict the position of five lobes as they are anatomically connected. Fig.1 shows the trachea and five bronchi.

## *3.4. Loss function*

The model's loss function is an adapted version of the Dice loss, which measures the Dice loss function per class and then adds them. Measuring the Dice score individually on each class $c \in C$, such that $C$ is the set of all classes we want to segment, is essential as we want to take into consideration the massive class imbalance. In equation 1, the ground truth label for each voxel $i \in N$ is represented through the binary vector $g_{ic}$ whose size is the same as the number of classes. The predicted probability for voxel $i$ to belong to all of the $C$ classes is denoted as $p_{ic}$. The $\delta$ provides numerical stability by preventing a division by zero.

$$L = \sum_{c=1}^{C} \frac{\sum_{i=1}^{N} p_{ic} g_{ic} + \delta}{\sum_{i=1}^{N}(p_{ic})^2 + \sum_{i=1}^{N}(g_{ic})^2 + \delta} \quad (1)$$

The model's overall loss function combines the loss function at each output in the layer (Equation 2). Each loss function in the output layer has a weight λ1, λ2 that correspond to main and auxiliary output, determining the importance in the overall function.

$$Loss = \lambda_1 L_{main} + \lambda_2 L_{aux} \quad (2)$$

## 4. RESULTS

### *4.1. Segmentation results*

We set the learning rate at the beginning to 0.01 and reduced it by a factor of 10 if there was no improvement in the validation score for 50 epochs; each epoch comprised the 50 images randomly augmented. We set the dropout to 0.5 and weights were set to $\lambda_1, \lambda_2$ = [0.5,0.5]. The model took 35h to train on a 24GB NVIDIA TITAN RTX. To evaluate the performance, we compared our model to that of others; not only other state-of-the-art models but also models incorporating the MTL component vs not incorporating it. Evaluating our model's performance, it was compared against the following (see Table III): a) the original deep convolutional neural network (DCNN) applied to the LUNA16 data set, b) state-of-the-art PLS-Net *(*Lee et al., 2019), and c) a V-Net without MTL. Comparing both models with and without MTL is important as most lobar segmentation studies used different datasets, and the specific dataset in our case influenced our decision to incorporate the trachea and bronchi into our segmentation task. Our specific dataset allowed us to incorporate the trachea and bronchi into segmentation.

TABLE III
COMPARISON: DICE SCORES OF THE DIFFERENT MODELS ON NOMAL LUNG STRUCTURE GIVEN IN THE FORM (MEAN ± STANDARD DEVIATION), AN ASTERISK SHOWS A SIGNIFICANT DIFFERENCE ($p<0.05$) COMPARED WITH V-NET

| Class | Dice Score | | | |
|---|---|---|---|---|
| | *V-Net with MTL* | *V-Net* | *DCNN* (Tang et al., 2019) [a] | *PLS-Net* (Lee et al., 2019) |
| LR lobe | 0.962±0.041* | 0.938±0.061 | 0.925 | 0.963±0.043 |
| MR lobe | 0.931±0.064* | 0.916±0.088 | 0.806 | 0.936±0.059 |
| UR lobe | 0.952±0.047* | 0.938±0.032 | 0.930 | 0.962±0.045 |
| LL lobe | 0.967±0.043* | 0.952±0.037 | 0.953 | 0.958±0.045 |
| UL lobe | 0.969±0.046* | 0.955±0.054 | 0.961 | 0.961±0.031 |
| Trachea [b] | 0.972±0.052 | - | - | - |
| Bronchi [b] | 0.649±0.123 | - | - | - |

[a] DCNN performed segmentation on the same data (LUNA16).
[b] Other models did not segment the trachea and bronchi.

Compared to the published models (Tang et al., 2019, Lee et al., 2019), our results achieve similar state-of-the-art accuracy. PLS-Net was trained on a non-publicly available dataset of 210 CT scans; this dataset is larger than any other previous study. While it is unclear which factors had contributed to the PLS-Net model achieving the high accuracy, it is appreciated that underlying performance of deep learning models is strongly linked to the size of the dataset (Litjens et al., 2017). Our model's good performance can be attributed to the availability of new recently released dataset (50 segmented CT scans) and by incorporating MTL into the network. Furthermore, the segmentation of the trachea was successfully incorporated without sacrificing the performance of lobar segmentation. The underperformance of the lobar and sublobar airways could be attributed to various reasons including smaller volume; these values were consistent with the training performance, thus we can argue that our model didn't overfit but rather wasn't able to focus on these volumes.

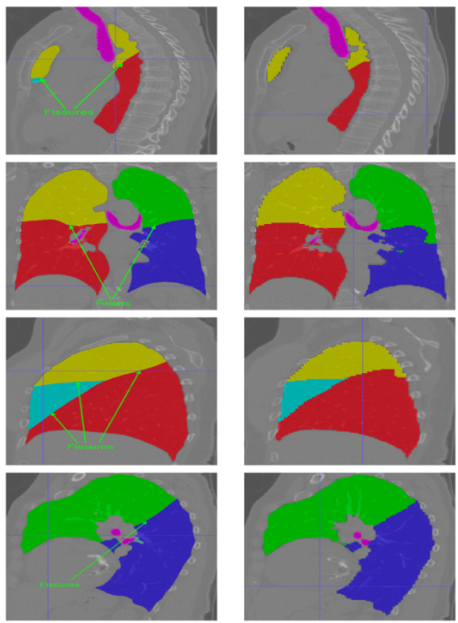

(a) Reference truth     (b) V-Net with MTL

Fig. 3. Visualisation: comparing the ground truth segmentation vs the segmentation results obtained from V-Net and MTL; the fissures are seen on the ground truth image.

## 4.2. Results of segmentation in patients with lung disease

Our model segmented a total of 24 thoracic CT scans, each from a patient with lung disease. The results are shown in Table III -- each ROI's performance separately with the small number of patients. As none of the CT scans held patients with such diseases through the training, a slight decrease in the Dice score is seen when comparing the values (see Table II). A condition that affected the segmentation the most was collapsed lung, with a mean drop of 0.032 in Dice score. In contrast, the disease that affected it most trivially was cancer, with a mean drop of 0.019. Notably, the lobar and sublobar airway segmentation suffered the most significant decrease in accuracy within the collapsed lung. It dropped 0.091, most likely due to the effect of loss of airway conspicuity.

TABLE IV

COMPARISON: DICE SCORES OF PATIENTS WITH DISEASES (COVID-19, cancer, collapsed lung and COPD) bold font are results from V-Net with MTL, the regular font is result from normal V-Net.

| | Cancer | | | | | | |
|---|---|---|---|---|---|---|---|
| Class | Patient 1 | Patient 2 | Patient 3 | Patient 4 | Patient 5 | Patient 6 | Mean |
| LR lobe | **0.959** | **0.952** | **0.954** | **0.974** | **0.986** | **0.960** | **0.964** |
| | 0.913 | 0.917 | 0.901 | 0.908 | 0.911 | 0.922 | 0.912 |
| MR lobe | **0.913** | **0.927** | **0.913** | **0.897** | **0.900** | **0.937** | **0.914** |
| | 0.869 | 0.863 | 0.865 | 0.863 | 0.887 | 0.874 | 0.870 |
| UR lobe | **0.911** | **0.945** | **0.871** | **0.947** | **0.881** | **0.950** | **0.917** |
| | 0.896 | 0.903 | 0.853 | 0.889 | 0.893 | 0.901 | 0.889 |
| LL lobe | **0.956** | **0.977** | **0.958** | **0.956** | **0.926** | **0.934** | **0.951** |
| | 0.923 | 0.951 | 0.945 | 0.934 | 0.916 | 0.932 | 0.933 |
| UL lobe | **0.958** | **0.974** | **0.953** | **0.961** | **0.926** | **0.922** | **0.949** |
| | 0.912 | 0.945 | 0.934 | 0.939 | 0.912 | 0.931 | 0.929 |
| Trachea | **0.968** | **0.979** | **0.972** | **0.965** | **0.958** | **0.971** | **0.969** |
| Bronchi | **0.624** | **0.633** | **0.619** | **0.646** | **0.624** | **0.635** | **0.630** |

| | COVID-19 | | | | | | |
|---|---|---|---|---|---|---|---|
| Class | Patient 1 | Patient 2 | Patient 3 | Patient 4 | Patient 5 | Patient 6 | Mean |
| LR lobe | **0.949** | **0.971** | **0.935** | **0.946** | **0.916** | **0.967** | **0.947** |
| | 0.910 | 0.925 | 0.905 | 0.911 | 0.901 | 0.923 | 0.912 |
| MR lobe | **0.90** | **0.922** | **0.923** | **0.883** | **0.921** | **0.880** | **0.905** |
| | 0.854 | 0.873 | 0.861 | 0.865 | 0.852 | 0.874 | 0.863 |
| UR lobe | **0.883** | **0.942** | **0.942** | **0.914** | **0.870** | **0.933** | **0.914** |
| | 0.868 | 0.875 | 0.892 | 0.871 | 0.889 | 0.874 | 0.878 |
| LL lobe | **0.950** | **0.970** | **0.949** | **0.944** | **0.941** | **0.967** | **0.953** |
| | 0.923 | 0.915 | 0.924 | 0.928 | 0.933 | 0.937 | 0.927 |
| UL lobe | **0.978** | **0.973** | **0.949** | **0.935** | **0.956** | **0.975** | **0.961** |
| | 0.953 | 0.946 | 0.942 | 0.914 | 0.944 | 0.938 | 0.939 |
| Trachea | **0.973** | **0.968** | **0.952** | **0.961** | **0.967** | **0.958** | **0.963** |
| Bronchi | **0.645** | **0.612** | **0.674** | **0.611** | **0.638** | **0.623** | **0.633** |

| | COPD | | | | | | |
|---|---|---|---|---|---|---|---|
| Class | Patient 1 | Patient 2 | Patient 3 | Patient 4 | Patient 5 | Patient 6 | Mean |
| LR lobe | **0.944** | **0.96** | **0.969** | **0.929** | **0.961** | **0.962** | **0.954** |
| | 0.911 | 0.944 | 0.892 | 0.896 | 0.867 | 0.919 | **0.905** |
| MR lobe | **0.929** | **0.88** | **0.939** | **0.861** | **0.866** | **0.932** | **0.901** |
| | 0.904 | 0.823 | 0.901 | 0.835 | 0.814 | 0.895 | **0.861** |
| UR lobe | **0.963** | **0.944** | **0.970** | **0.934** | **0.894** | **0.945** | **0.941** |
| | 0.958 | 0.893 | 0.922 | 0.911 | 0.870 | 0.926 | **0.913** |
| LL lobe | **0.977** | **0.973** | **0.944** | **0.941** | **0.963** | **0.961** | **0.960** |
| | 0.958 | 0.973 | 0.918 | 0.924 | 0.955 | 0.938 | **0.944** |
| UL lobe | **0.984** | **0.964** | **0.950** | **0.950** | **0.967** | **0.965** | **0.963** |
| | 0.964 | 0.908 | 0.937 | 0.946 | 0.950 | 0.913 | **0.936** |
| Trachea | **0.953** | **0.946** | **0.974** | **0.963** | **0.943** | **0.969** | **0.958** |
| Bronchi | **0.642** | **0.546** | **0.634** | **0.641** | **0.654** | **0.652** | **0.627** |

| | Collapsed Lung | | | | | | |
|---|---|---|---|---|---|---|---|
| Class | Patient 1 | Patient 2 | Patient 3 | Patient 4 | Patient 5 | Patient 6 | Mean |
| LR lobe | **0.951** | **0.945** | **0.964** | **0.955** | **0.934** | **0.961** | **0.952** |
| | 0.887 | 0.913 | 0.892 | 0.896 | 0.867 | 0.919 | 0.896 |
| MR lobe | **0.834** | **0.912** | **0.960** | **0.888** | **0.893** | **0.902** | **0.897** |
| | 0.785 | 0.863 | 0.823 | 0.806 | 0.791 | 0.832 | 0.816 |
| UR lobe | **0.961** | **0.946** | **0.836** | **0.785** | **0.844** | **0.963** | **0.886** |
| | 0.901 | 0.893 | 0.767 | 0.712 | 0.753 | 0.885 | 0.815 |
| LL lobe | **0.941** | **0.944** | **0.951** | **0.935** | **0.910** | **0.963** | **0.941** |
| | 0.893 | 0.904 | 0.889 | 0.911 | 0.892 | 0.913 | 0.900 |
| UL lobe | **0.956** | **0.958** | **0.944** | **0.940** | **0.911** | **0.954** | **0.944** |
| | 0.899 | 0.908 | 0.902 | 0.905 | 0.867 | 0.907 | 0.898 |
| Trachea | **0.966** | **0.963** | **0.952** | **0.954** | **0.958** | **0.949** | **0.957** |
| Bronchi | **0.583** | **0.546** | **0.544** | **0.567** | **0.553** | **0.555** | **0.558** |

Importantly, the comparison of Table IV results vs Table III confirms that MTL V-Net (disease patients) outperforms the single-use V-Net model (healthy patients), and further confirms that the trachea and bronchi segmentation allows a more robust model giving it useful contextual intelligence from the viable CT data in the context of disease affecting lung anatomy.

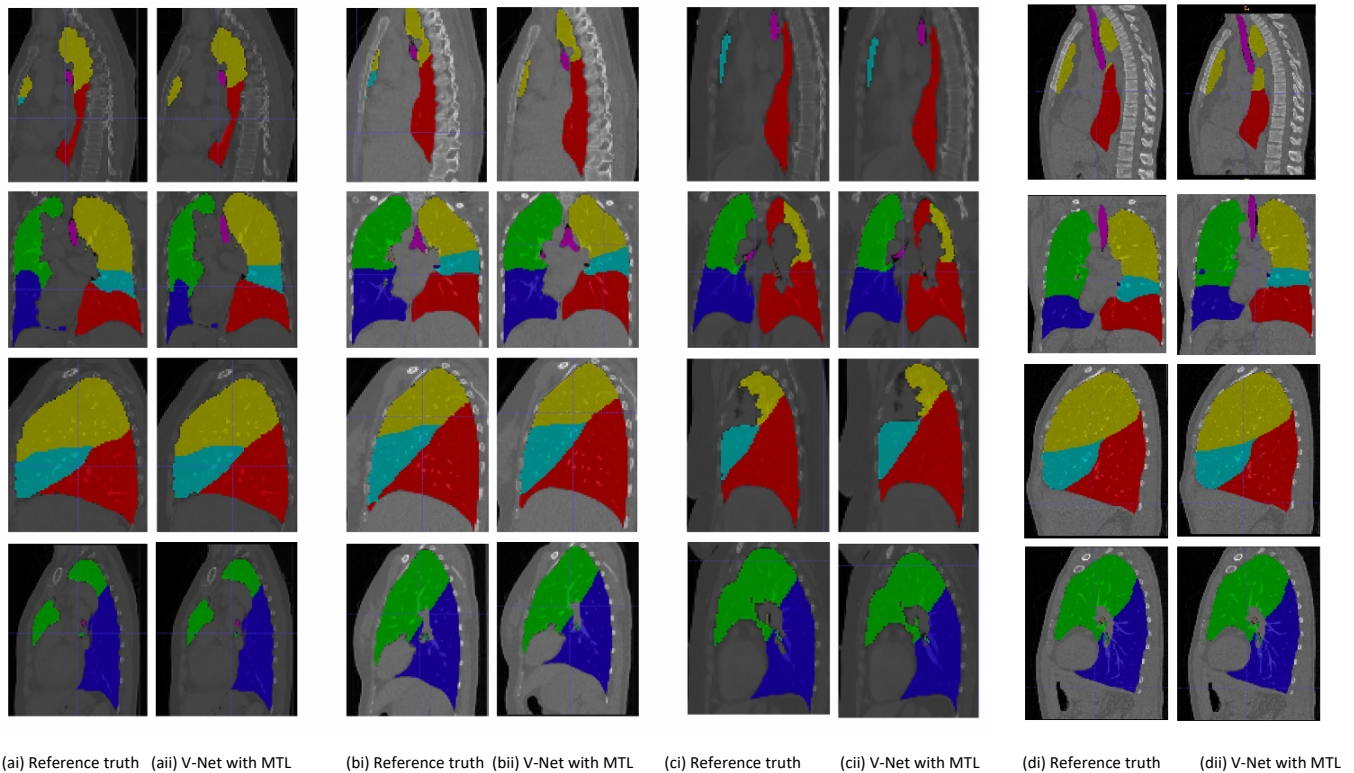

(ai) Reference truth  (aii) V-Net with MTL    (bi) Reference truth  (bii) V-Net with MTL    (ci) Reference truth   (cii) V-Net with MTL    (di) Reference truth    (dii) V-Net with MTL

Fig. 4. Visualisation: comparing the ground truth segmentation versus the segmentation results obtained from V-Net and MTL; for Cancer (a), COVID-19 pneumonitis (b), collapsed lung (c), and COPD (d).

5. DISCUSSION

With no current deep learning model able to effectively recognise anatomical variations and quantify lobar disease distribution on CT images, an automated method equivalent to visual perception of lobar anatomy may still not be feasible for a long time. Yet, there should be a clear benefit of making an incremental step towards this goal. What sets apart our model is the adeptness to deal with more challenging cases -- when increased lung attenuation on CT images makes it difficult to accurately delineate pulmonary lobes. While reflecting on the study's outcomes, several important details are worth highlighting.

First, the model achieved near state-of-the-art results despite being trained on a relatively small dataset, which is important as the scarcity of labelled training data in healthcare is regarded as an unaccommodating issue. Another detail is that the training data did not include any severe lung disease with marked differences in lung attenuation. Nonetheless, model robustness was proved when later being validated on the external dataset, with extensive lung abnormalities. And lastly, to our awareness, this is the first work that validates lobar segmentation in the context of structural abnormalities caused by specific lung conditions (cancer, viral pneumonitis, lobar collapse) affecting lung attenuation.

Answering the problem of lobar quantification, we explored whether processing of two interrelated pulmonary tissues together, using deep learning (MTL) improves the segmentation and makes it more robust to anatomical differences in the lungs. Our principal findings firmly mark that simultaneous segmentation of anatomically relevant tissues was advantageous to the task as mentioned above. The results may hold some important implications for the field. Firstly, our model may be useful to solve clinical tasks with limited data. Secondly, new avenues linked to tissues/organs interrelatedness may be explored to achieve more reliable image segmentation tasks. At the same time, we are acknowledging that unlike the lobar task, the auxiliary element has ample room for potential improvement. Yet, it was not the purpose of this study to maximise the accuracy of the lobar bronchi segmentation. Our main intention was to investigate if using the latter as an auxiliary task would improve the main task. This aim partly justifies somewhat the underperformance in the auxiliary part. Hence, tackling those as equal and optimizing the network to the auxiliary and main task will most likely improve the performance on the segmentation of the lobar bronchi.

Comparing the V-Net MTL performance to PLS-Net see (Table II), no statistical difference was observed. However, a direct comparison is difficult: PLS-Net was trained on a different dataset, not publicly available. Given a lack of statistical significance, we cannot conclude that either model is statistically superior. Still, as it was trained on a dataset two times smaller, we believe that this is an advantage of V-Net MTL: medical datasets tend to be small, and the field welcomes models capable of performing in the conditions when data are scarce.

The lung disease cases show that our model remains robust when dealing with deformed fissure shapes or lung attenuation abnormalities. As such, it only suffers a slight drop in performance. Equally important, the performance drop of V-NET with MTL is much smaller than the drop in V-Net alone, leading us to believe that the airway segmentation indeed helps the model segment the lobes.

Lastly, in the case of lobar segmentation, our auxiliary task only included lobar bronchi and trachea. We did not examine segmentation of other interrelated pulmonary tissues whose auxiliary role in the V-Net-MTL architecture might be even more helpful as seen in other studies where MTL was used to improve the segmentation of multiple organs at risk (He et al., 2020). In future studies, it would be advisable to consider other vital respiratory areas and incorporate them into the proposed MTL framework – each of these need to be carefully segmented.

6. CONCLUSION

The study explored the beneficial idea of utilising tracheobronchial airway background information in the segmentation of pulmonary lobes. By itself, our approach demonstrated an improved ability of the AI-enabled model to recognise lobar lung tissue in the context of abnormalities present on CT scans of various conditions such as COPD, viral pneumonitis, cancer or significant lobar collapse, with imaging manifestations including distortion of pulmonary fissures and lung parenchyma, emphysema and lobar collapse with loss of airway aeration. The system's statistical power bypasses the necessity of handcrafted feature engineering. It automatically delineates lung lobes with the fissural boundaries, useful for diagnosis and targeted preoperative planning. The proposed model can be readily adopted in an existing clinical setting as a useful clinical tool for radiologists and researchers. We assert that the proposed approach could support novel computational systems to more accurately define lobar distribution of lung disease.

REFERENCES

Armato III, Samuel G.; Hadjiiski, Lubomir; Tourassi, Georgia D.; Drukker, Karen; Giger, Maryellen L.; Li, Feng; Redmond, George; Farahani, Keyvan; Kirby, Justin S.; Clarke, Laurence P. (2015). SPIE-AAPM-NCI Lung Nodule Classification Challenge Dataset. The Cancer Imaging Archive. https://doi.org/10.7937/K9/TCIA.2015.UZLSU3FL

Bian Y, Guo S, Jiang H et al (2019) Relationship between radiomics and risk of lymph node metastasis in pancreatic ductal adenocarcinoma. Pancreas 48:1195–1203

Caruana, R., 1997. Multitask learning. Machine learning, 28(1), pp.41-75.

Clark K, Vendt B, Smith K, et al. The Cancer Imaging Archive (TCIA): Maintaining and Operating a Public Information Repository. Journal of Digital Imaging. 2013; 26(6): 1045-1057. doi: 10.1007/s10278-013-9622-7.

Doel, T., Gavaghan, D.J. and Grau, V., 2015. Review of automatic pulmonary lobe segmentation methods from CT. Computerized Medical Imaging and Graphics, 40, pp.13-29.

Doel T. Pulmonary Toolkit. https://github.com/tomdoel/pulmonarytoolkit. Accessed Apr 10, 2021

Fang, X., Li, X., Bian, Y., Ji, X. and Lu, J., 2020. Radiomics nomogram for the prediction of 2019 novel coronavirus pneumonia caused by SARS-CoV-2. European radiology, pp.1-14.

Fedorov A., Beichel R., Kalpathy-Cramer J., Finet J., Fillion-Robin J-C., Pujol S., Bauer C., Jennings D., Fennessy F.M., Sonka M., Buatti J., Aylward S.R., Miller J.V., Pieper S., Kikinis R. 3D Slicer as an Image Computing Platform for the Quantitative Imaging Network. Magn Reson Imaging. 2012 Nov;30(9):1323-41. PMID: 22770690. PMCID: PMC3466397.

Ferreira, F.T., Sousa, P., Galdran, A., Sousa, M.R. and Campilho, A., 2018, July. End-to-end supervised lung lobe segmentation. In 2018 International Joint Conference on Neural Networks (IJCNN) (pp. 1-8). IEEE.

Gevenois PA, de Maertelaer V, De Vuyst P, Zanen J, Yernault JC. Comparison of computed density and macroscopic morphometry in pulmonary emphysema. Am J Respir Crit Care Med. 1995 Aug;152(2):653-7. doi: 10.1164/ajrccm.152.2.7633722. PMID: 7633722.

Global Strategy for the Diagnosis, Management and Prevention of COPD, Global Initiative for Chronic Obstructive Lung Disease (GOLD). 2018 revision. Available from: http://www.goldcopd.org (accessed 4th May, 2021).

Goncharov, M., Pisov, M., Shevtsov, A., Shirokikh, B., Kurmukov, A., Blokhin, I., Chernina, V., Solovev, A., Gombolevskiy, V., Morozov, S. and Belyaev, M., 2020. CT-based COVID-19 Triage: Deep Multitask Learning Improves Joint Identification and Severity Quantification. arXiv preprint arXiv:2006.01441.

Hancock, M.C. and Magnan, J.F., 2016. Lung nodule malignancy classification using only radiologist-quantified image features as inputs to statistical learning algorithms: probing the Lung Image Database Consortium dataset with two statistical learning methods. Journal of Medical Imaging, 3(4), p.044504.

He, T., Guo, J., Wang, J., Xu, X. and Yi, Z., 2019. Multi-task Learning for the Segmentation of Thoracic Organs at Risk in CT images. In SegTHOR@ ISBI.

He, T., Hu, J., Song, Y., Guo, J. and Yi, Z., 2020. Multi-task learning for the segmentation of organs at risk with label dependence. Medical Image Analysis, 61, p.101666.

Kuhnigk JM, Dicken V, Zidowitz S, Bornemann L, Kuemmerlen B, Krass S, Peitgen HO, Yuval S, Jend HH, Rau WS, Achenbach T. Informatics in radiology (infoRAD): new tools for computer assistance in thoracic CT. Part 1. Functional analysis of lungs, lung lobes, and bronchopulmonary segments. Radiographics. 2005 Mar-Apr;25(2):525-36. doi: 10.1148/rg.252045070. PMID: 15798068.

Lee H, Matin T, Gleeson F, Grau V (2019) Efficient 3D fully con-volutional networks for pulmonary lobe segmentation in ct images.arXiv:1909.07474

Litjens, G., Kooi, T., Bejnordi, B.E., Setio, A.A.A., Ciompi, F., Ghafoorian, M., Van Der Laak, J.A., Van Ginneken, B. and Sánchez, C.I., 2017. A survey on deep learning in medical image analysis. Medical image analysis, 42, pp.60-88.

Lu, H., Arshad, M., Thornton, A., Avesani, G., Cunnea, P., Curry, E., Kanavati, F., Liang, J., Nixon, K., Williams, S.T. and Hassan, M.A., 2019. A mathematical-descriptor of tumor-mesoscopic-structure from computed-tomography images annotates prognostic-and molecular-phenotypes of epithelial ovarian cancer. *Nature communications*, *10*(1), pp.1-11.

Mansoor, A., Bagci, U., Foster, B., Xu, Z., Papadakis, G.Z., Folio, L.R., Udupa, J.K. and Mollura, D.J., 2015. Segmentation and image analysis of abnormal lungs at CT: current approaches, challenges, and future trends. RadioGraphics, 35(4), pp.1056-1076.

Milletari, F., Navab, N. and Ahmadi, S.A., 2016, October. V-net: Fully convolutional neural networks for volumetric medical image segmentation. In 2016 fourth international conference on 3D vision (3DV) (pp. 565-571). IEEE.

Morozov, S.P., Andreychenko, A.E., Pavlov, N.A., Vladzymyrskyy, A.V., Ledikhova, N.V., Gombolevskiy, V.A., Blokhin, I.A., Gelezhe, P.B., Gonchar, A.V. and Chernina, V.Y., 2020. MosMedData: Chest CT Scans With COVID-19 Related Findings Dataset. arXiv preprint arXiv:2005.06465.

Ng MY, Lee EYP, Yang J, Yang F, Li X, Wang H, Lui MM, Lo CS, Leung B, Khong PL, Hui CK, Yuen KY, Kuo MD. Imaging Profile of the COVID-19 Infection: Radiologic Findings and Literature Review. Radiol Cardiothorac Imaging. 2020 Feb 13.

Park, J., Yun, J., Kim, N., Park, B., Cho, Y., Park, H. J., Song, M., Lee, M., & Seo, J. B. (2020). Fully Automated Lung Lobe Segmentation in Volumetric Chest CT with 3D U-Net: Validation with Intra- and Extra-Datasets. Journal of digital imaging, 33(1), 221–230. https://doi.org/10.1007/s10278-019-00223-1

Ruder, S., 2017. An overview of multi-task learning in deep neural networks. *arXiv preprint arXiv:1706.05098*.

Schlemper, J., Oktay, O., Schaap, M., Heinrich, M., Kainz, B., Glocker, B. and Rueckert, D., 2019. Attention gated networks: Learning to leverage salient regions in medical images. Medical image analysis, 53, pp.197-207

Shelhamer E, Long J, Darrell T. Fully Convolutional Networks for Semantic Segmentation. IEEE Trans Pattern Anal Mach Intell. 2017 Apr;39(4):640-651. doi: 10.1109/TPAMI.2016.2572683. Epub 2016 May 24. PMID: 27244717.

Shen, D., Wu, G. and Suk, H.I., 2017. Deep learning in medical image analysis. Annual review of biomedical engineering, 19, pp.221-248.

Tang, H., Zhang, C. and Xie, X., 2019, April. Automatic Pulmonary Lobe Segmentation Using Deep Learning. In 2019 IEEE 16th International Symposium on Biomedical Imaging (ISBI 2019) (pp. 1225-1228). IEEE.

Wang, C., 2017, June. Segmentation of multiple structures in chest radiographs using multi-task fully convolutional networks. In Scandinavian Conference on Image Analysis (pp. 282-289). Springer, Cham.